# Linear measurements from nonlinear sensors: identifying distortion with incidental noise

Lachlan J. Gunn, Andrew Allison, and Derek Abbott

lachlan@gunn.ee    { andrew.allison, derek.abbott }@adelaide.edu.au

Aalto University
*Finland*

The University of Adelaide
*Australia*

Nonlinearity in many systems is heavily dependent on component variation and environmental factors such as temperature. This is often overcome by keeping signals close enough to the device's operating point that it appears approximately linear.

But as the signal being measured becomes larger, the deviation from linearity increases, and the device's nonlinearity specification will be exceeded. This limits the range over which the device will produce directly useful measurements, often to far less than the device's safe range of operation.

But though these nonlinear measurements may exceed the device's nonlinearity specification, they nevertheless contain useful information. We have shown that the input noise of a system can provide the necessary information to determine the function relating the input signal to the output signal, allowing us to invert it numerically and recover the original signal, even if the device has been driven quite heavily into nonlinearity.

In this article, we will provide an overview the technique and results from [1]–[3], and show how it can be used to provide real-time compensation of distortion, providing a significant advantage over the state-of-the-art techniques standardized in [4] when dealing with grossly-distorting systems. We begin by showing how noise can be used to measure the nonlinearity of a system; then we discuss different models of nonlinearity that are suitable for use in real-time compensation tools, and finally we show how this can be applied in practice.

## Measuring nonlinearity

The first step is to *measure* the nonlinearity, in order that we can compensate for it; this is a system identification problem that has been tackled in many different ways depending on the application. In our work, we follow the lead of the data-converter literature, mapping each digitized value (or *code*) onto its corresponding range of input values, and aiming to ensure that these ranges are as similar as possible. The difference between the width of these ranges and the average width is known as the *differential*

---

*nonlinearity* (DNL) [4]. Integrating these differential nonlinearities yields the *integral nonlinearity* (INL) [4], the variation between the measured value at some input level and a linear fit between the limits of the input range.

Directly measuring the widths of these ranges is slow and expensive, requiring very high-precision DC sources. The current state of the art is standardized in [4, §4.7]; these methods require a DC source with a resolution and accuracy at least four times that of the digitizer being measured; the source is adjusted until it lies on the boundary between each pair of adjacent codes.

An alternative approach is statistical in nature: apply a signal to the measurement system that has a known distribution, then measure the distribution of the measurements: codes whose ranges are wider or narrower will be obtained more or less frequently than the input signal distribution would predict. Analysis is simplest with a uniformly-distributed input signal, such as a triangle function; however, these are very difficult to generate with high precision. Sinusoidal signals can be generated with excellent levels of spectral purity, resulting in a highly-precise input signal [5]. The two approaches can be combined as in [5], speeding the measurement process without high-precision equipment.

These approaches work very well for factory testing, but in the field they have a major drawback: the device under test must be disconnected from its input and excited with a high-precision test signal. This introduces dead time, and requires that each device contain a highly precise—and expensive—signal source.

But this is not necessary: electronic devices provide their own test signal in the form of Gaussian noise, and the signal being measured continuously sweeps this test signal across the whole range of interest. Non-electronic process noise—e.g. vibration of an object whose position is being measured—can serve the same purpose if it remains similarly constant over time and the measurement system has sufficient bandwidth and a short-enough averaging time.

Suppose the system has a static nonlinear response $y = f(x)$, with band-limited input signal $x(t)$ and white Gaussian input noise $n(t)$. This results in an output signal of

$$y(t) = f\big(x(t) + n(t)\big).$$

Our goal is to find a *compensating function* $g(x)$ such that $g(y(t)) \propto x(t)$; we will then use this to build a device that takes a distorted input signal, to produce an output signal that approximates the distortion-free input signal.

Where the noise is small relative to the signal, $f(x)$ can be approximated by its first-order Taylor series truncation around $x(t)$

$$y(t) \approx f\big(x(t)\big) + n(t) f'\big(x(t)\big).$$

When the derivative $f'(x(t))$ — the differential gain of the system — is large, this results in a larger noise signal at the output, as illustrated in Figure 1. It is this effect that we will use to identify $f$.

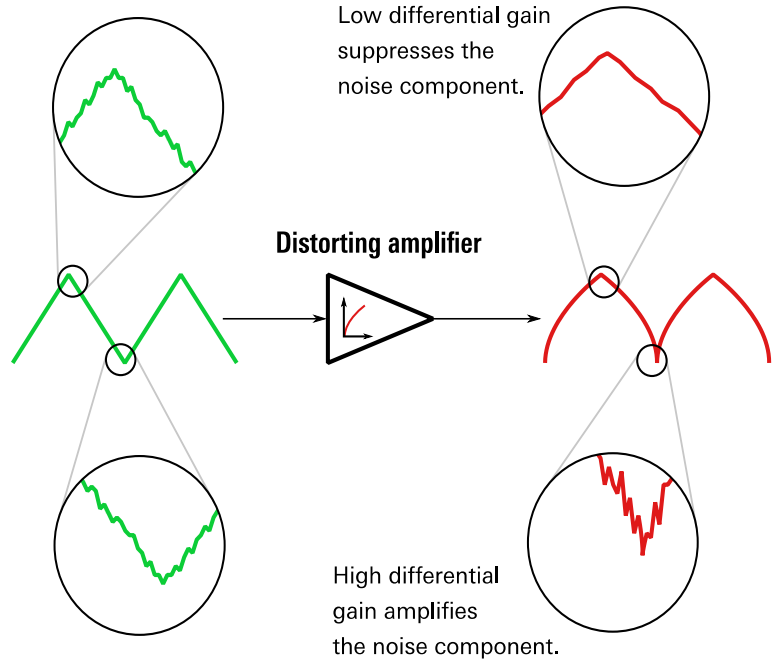

Figure 1: The effect of distortion on a noisy signal. In regions of high differential gain, the noise is amplified; conversely, when the differential gain is low, this results in suppression of the signal's noise component.

As the input signal $x(t)$ is band-limited, the distorted signal component $f(x(t))$ and noise component $n(t)f'(x(t))$ can be isolated by low-pass and high-pass filtering, respectively. As $x(t)$ varies, so will the amplitude of the noise component: if the noise at the input has standard deviation $\sigma_i$, then the noise at the output will have standard deviation $\sigma_o(t) = |f'(x(t))|\sigma_i$. If the sign of $f'(x(t))$ is positive—i.e. $f$ is monotonically increasing—over the region of excitation then $f$ can be obtained and inverted by numerical integration; the difference between the input signal at time $t_0$ and at time $t_n$ is given by:

$$x(t_n) - x(t_0) = \int_{f(x(t_0))}^{f(x(t_n))} \frac{\sigma_i}{\sigma_o\left(f(x(\tau))\right)} df(x(t)).$$

This demonstrates that it is possible, in principle, to identify the nonlinearity of a nonlinear system from its output only, provided the bandwidth of the signal of interest is small enough relative to the noise bandwidth. However, this approach has several drawbacks: first, the noise bandwidth must be quite large, as power from the signal that bleeds into the noise measurement band will tend to bias the signal; this problem is exacerbated when the noise power is low. This source of bias makes noise measurement at the compensator input problematic, as nonlinear distortion tends to add to the introduce high-frequency harmonic content. Additionally, methods based on numerical integration are inefficient; we must consider other approaches if we are to perform such compensation in real time.

## Representing nonlinearity

If the approach is to be used in real-time, we must drastically reduce its computational burden. This means that we cannot look at each prior measurement, but must instead reduce the raw measurements to a more compact representation of the function that can be continuously updated.

The representation has three main requirements. The first of these is practical:

1. **Efficiency.** It must be possible to efficiently compute and update the representation. In practice, this means that numerical integration must be avoided.

The remaining two follow from a spirit of "first, do no harm":

2. **Representation of linear functions.** The representation must be capable of exactly modelling a linear system.
3. **Differentiability.** The representation must not itself introduce significant harmonic content. We therefore avoid jumps and sharp corners, preferring differentiable approximations.

Table 1: Nonlinear transfer function representations. Code: ◎ = false, ◐ = partially true, ● = true.

|  | Efficient evaluation | Efficient update | Perfect representation of linear functions | Differentiability |
|---|---|---|---|---|
| Polynomial (Legendre, etc.) | ◐ | ◐ | ● | ● |
| Fourier | ◎ | ◐ | ◎ | ● |
| Piecewise-linear | ● | ● | ● | ◎ |
| Integrated piecewise-linear | ◕ | ● | ● | ● |

Several possible representations are shown in Table 1. We considered global representations such as polynomials, but these have the disadvantage of requiring a relatively large number of computations to compute and update; tolerating more complex nonlinearities requires an increase in degree that slows down evaluation across the entire input range. Using a Fourier basis rather than orthogonal polynomials has similar downsides, but also suffers from the Gibbs effect and so also has difficulty representing linear functions.

Piecewise representations, on the other hand, can be much more efficient: rather than increasing the degree of the approximating polynomials, thereby slowing down every operation, it is possible to simply split the input range into smaller pieces, costing memory but not time when it is evaluated.

A piecewise linear representation can be computed efficiently whilst still being able to represent linear systems, but introduces sharp corners into the signals being processed, potentially modifying the spectral characteristics of the original signal.

An alternative is to represent the compensation function as the *integral* of a continuous piecewise-linear function. This can be efficiently evaluated without resort to numerical integration, making this representation almost as efficient as a piecewise-linear one.

**Integrated piecewise-linear**

Our test platform is based around an STM32F4DISCOVERY evaluation board [6, p. 32], whose processor, the STM32F407VG contains a fast floating-point unit. We therefore opted initially to use an integrated piecewise-linear approximation to the compensation function, represented by triangular radial basis functions [7], as shown in Figure 2.

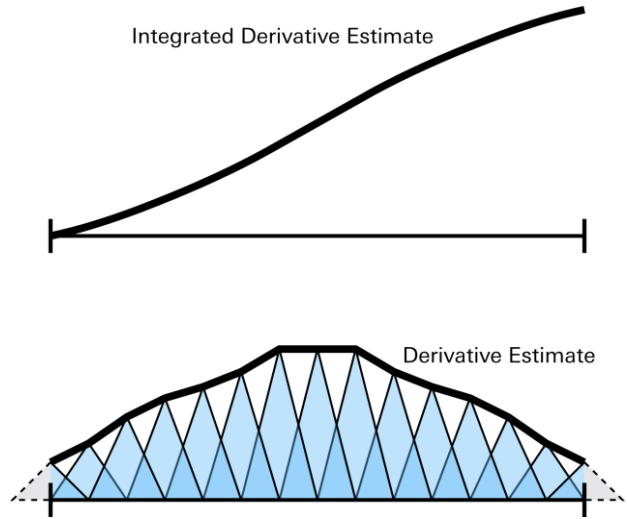

Figure 2: Integrated radial basis function model of the compensation function. By integrating a continuous representation of estimated inverse derivative as above, we avoid sharp "corners" in the compensated signal.

The estimated derivative $\widehat{f^{-1}}'(y)$ of the compensation function is therefore written as the sum of translated and scaled basis functions

$$r(x) = \begin{cases} 1 - \frac{|x|}{\Delta}, & if\,|x| < \Delta \\ 0, & \text{otherwise} \end{cases}$$

where $\Delta$ is the width of the basis function. We space these apart by $\Delta = (y_{\max} - y_{\min})/n_{\text{pieces}}$, such that almost every point in the system's output range is covered by two basis functions. This yields the model representation

$$\widehat{f^{-1}}'(y) = \sum_{k=0}^{n_{\text{pieces}}-1} c_k r(y - y_{\min} - k\Delta)$$

with integral

$$\int_{-\infty}^{f(x(t))} \widehat{f^{-1}}'(y)\,dy = \Delta \sum_{k=-\infty}^{n-1} c_k + \frac{1}{2}\Delta c_n + \big(f(x(t)) - n\Delta\big)c_n + \frac{\big(f(x(t)) - n\Delta\big)^2 (c_{n+1} - c_n)}{2\Delta},$$

where $n = \lfloor (y - y_{\min})/\Delta \rfloor$ and $c_k$ is the weight of the basis function centered at $y_{\min} + k\Delta$.

This is a piecewise-quadratic function; the coefficients for each piece can be precomputed when the model is updated, making evaluation very efficient.

## Real-time compensation of distortion

### Floating-point implementation

We have developed a distortion compensation library suitable for devices with a floating-point unit[1]. The library's compensation algorithm works as follows, also illustrated in Figure 3:

1. Use a second-order Butterworth low-pass filter to isolate the distorted signal and subtract it from from the noise,
2. For batches of samples, estimate the standard deviation of the noise signal across the measurement range,
    a. Update the basis weights for the compensation function derivative according to the measured standard deviation in that segment of the measurement range,
    b. Re-integrate the derivative to obtain new parameters for the compensation function
3. Apply the compensation function to the raw signal

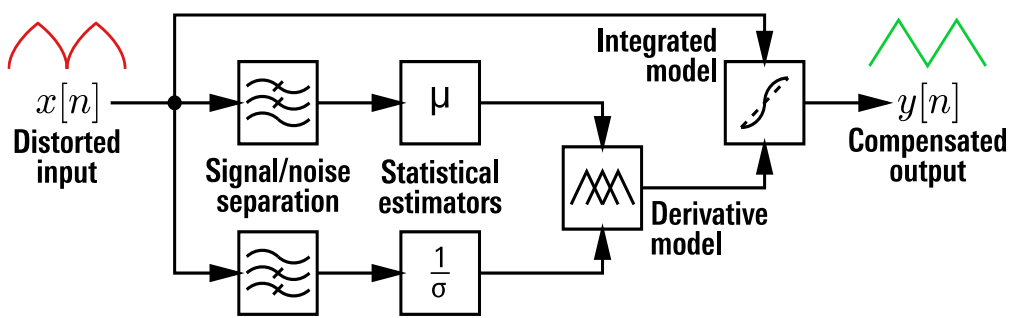

Figure 3: Distortion compensation library for floating-point implementation. The distorted input is processed to estimate the inverse function of the nonlinear response of the system that produced it. This compensation function is applied to the raw signal, yielding an estimate of the signal before it was distorted. Adapted from [2].

---

[1] Available at https://github.com/LachlanGunn/stochastic-instrumentation-tools

The basis weights are chosen to approximate the derivative of the inverse of the nonlinear response function of the system. From before, $\sigma_o(t) = |f'(x(t))|\sigma_i$; once again supposing that the $f$ is monotonically increasing, this implies that $\sigma_o(t)/\sigma_i = f'(x(t))$, and so we can estimate $\widehat{f^{-1}}'(y(t)) \propto 1/\hat{\sigma}_o(t)$. This is an expensive operation, requiring both computation of a square root and a division. We update the basis weights using a first-order IIR filter to average this estimate over many blocks, whilst allowing the weights to adapt to change in the nonlinear response of the system. We also take into account the value of the basis function at the point $y(t)$; basis weights are more strongly dependent on measurements taken near the center of the basis function. Full details are given in [2].

To evaluate this algorithm, we produced distorted signals by passing the output of a signal generator through a common-emitter amplifier without emitter degeneration, and then providing them to the on-chip ADC of the STM32F407 running our algorithm, writing the compensated signal to a DAC. The 10-bit ADC and 12-bit DAC each operate at a sample rate of 1.55 MHz. The mean and standard deviation of the signal are calculated from blocks of four samples, taken at 128-sample intervals. The integrated piecewise-linear model uses 256 triangular basis functions, which are reintegrated every 1024 blocks. The effect of the algorithm on a distorted triangle wave is shown in Figure 4.

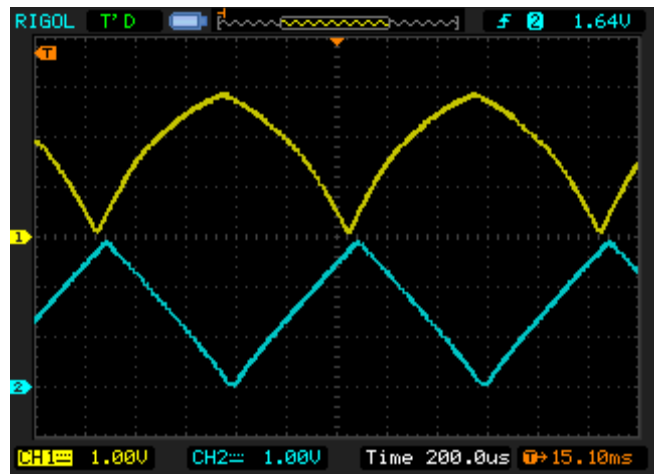

*Figure 4: Distorted triangle wave output by our overdriven common-emitter amplifier (channel 1; yellow), and the estimated pre-amplifier signal produced by our algorithm (channel 2; blue). The algorithm significantly reduces the level of distortion, making the triangular shape of the signal given to the amplifier clearly visible. Reproduced from [2].*

In order to quantify the algorithm's ability to reduce distortion, we used the same amplifier to distort sinusoidal signals at various amplitudes, as shown in Figure 5.

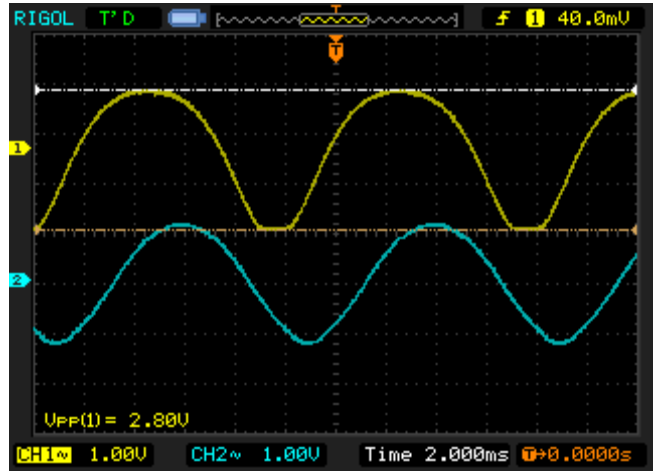

*Figure 5: Distorted sinusoid output by our overdriven common-emitter amplifier (channel 1; yellow), and the estimated pre-amplifier signal produced by our algorithm (channel 2; blue). The algorithm significantly reduces the level of distortion, even when the amplifier is clipping heavily. Reproduced from [2].*

The total harmonic distortion before and after applying our algorithm is shown in Figure 6. Our algorithm reduces the total harmonic distortion by between 10 dB and 15 dB at most levels, a significant improvement.

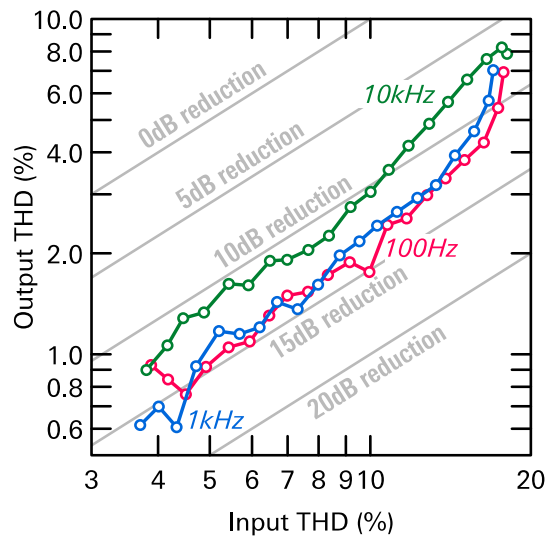

*Figure 6: Total harmonic distortion before and after compensation of a distorted sinusoid with the device presented in and as measured in [2], as a function of distortion level and signal frequency.*

However, this approach carries with it a large computational burden, and our implementation made heavy use of floating-point arithmetic, and in particular an expensive inverse-square-root operation; we are therefore motivated to find a more efficient approach to distortion compensation.

### Fixed-point compensation

Computing the derivative of the estimated inverse transformation requires a computationally expensive inverse-square-root operation. However, this and other floating-point operations can be avoided by the use of feedback instead of direct computation.

This is based on the observation that the noise distribution of a perfectly linear system will stay constant as the signal varies. When the noise of the compensated signal rises above its average value, this indicates that the differential gain at this signal level is too great, and must be suppressed, and vice-versa.

This suggests we might use negative feedback to set the model parameters: when the level of noise is greater than average, we reduce the differential gain, and when less than average we increase it. Because the noise is measured at the output of the compensator, the signal will be less likely to spill over into the noise band, as the compensator has already removed most of the distortion, eliminating some of the harmonic content. However, this is difficult with our previous model, since the integration is too slow to do for every block. Weight updates will occur only with some delay, raising stability concerns [8]. Instead, we need a model that can be updated in real-time in response to deviations from linearity. We chose to use a piecewise-linear compensating function, with each fixed-size range of input codes mapped to a corresponding range of output codes, as illustrated in Figure 7. Updating the differential gain of a section requires involves incrementing or decrementing the corresponding span.

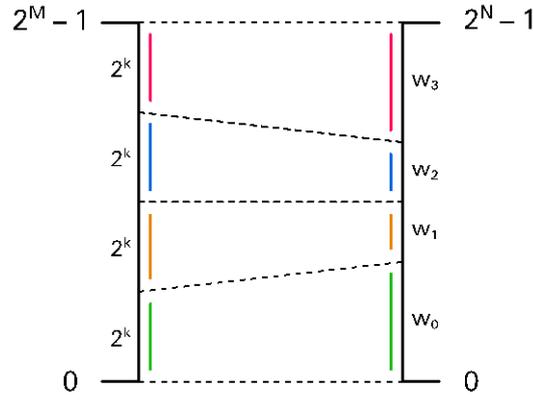

a) Partitioning of input and output spaces.

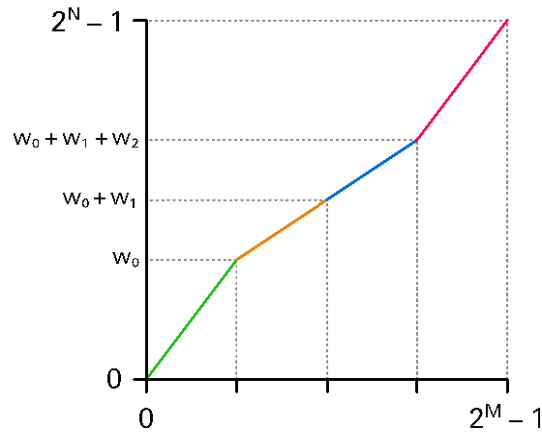

b) The resulting compensation function.

*Figure 7: A continuous piecewise-linear compensating function. The input space is partitioned into several ranges of equal width, whose span in the output space varies according to the parameters $w_i$. Adapted from [3].*

This is not difficult to evaluate using only fixed-point arithmetic. However, one difficulty still remains: the number of output codes is fixed, while the spans $w_i$ may vary. Whenever some $w_i$ is increased or decreased, the others must change accordingly. We reallocate this code in a fair manner by selecting at random another range to be widened or narrowed.

The feedback direction is determined by comparing the current noise variance with an average over the entire output range, calculated by low-pass filtering the noise variance measurements with a slow time-constant. We have evaluated this technique with simulated $\tanh(x)$-distorted data, yielding the results shown in Figure 8.

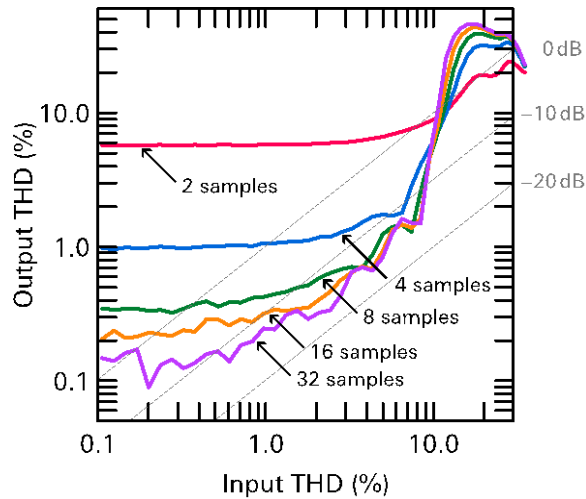

*Figure 8: Simulated total harmonic distortion achieved with feedback-based compensation. Improvements of 10–15 dB are achievable as before, but the output THD "bottoms out" when the input signal has little distortion. This noise floor depends on the number of samples per block (shown above), which determine the amount of noise present in the estimate of standard deviation. Adapted from [3].*

This approach avoids expensive inverse-square-root operations, but is more sensitive to sampling noise; as such, it is better suited to situations where distortion is large and the sampling rate is high enough to permit the use of large block sizes.

## Conclusions

In this article, we presented a new approach to the handling of static nonlinearity that can be applied in post-processing without any special knowledge of the function that has been applied to the signal. Nonlinearity induces a predictable change in the noise overlaid on the measured signal, and this variation can be used to infer the characteristics of the system's distortion. This opportunistic use of noise allows a degree of self-calibration is not possible with previous techniques, and will allow for the creation of flexible instrumentation that can maintain linearity over a far greater range of conditions.

## Biographies

**Lachlan J. Gunn** is a postdoctoral researcher in the Secure Systems Group at Aalto University in Finland. Prior to this, he was a research associate in the Computational Learning Systems laboratory of the University of South Australia. He received a Ph.D in Engineering from the University of Adelaide in 2018, and a B.Eng. (Hons.) and B.Math. & Comp.Sc. (Pure) degrees from The University of Adelaide, Australia, in 2012.

**Andrew Allison** received the B.Sc. degree in mathematical sciences and the B.Eng. (Hons.) degree in computer systems engineering from The University of Adelaide, in 1978 and 1995, respectively, and the Ph.D. degree in electrical and electronic engineering from The University of Adelaide in 2009. Since 1995, he has been with the School of Electrical and Electronic Engineering, University of Adelaide, as a Lecturer. Dr. Allison's research interests include probability, statistics and estimation, control theory, communication theory, and diffusion processes

**Derek Abbott** (M'85-SM'99-F'05) received the B.Sc. (Hons.) degree in physics from Loughborough University, Leicestershire, U.K., in 1982 and the Ph.D. degree in electrical and electronic engineering from The University of Adelaide, Australia. Since 1987, he has been with The University of Adelaide, where he is currently a full Professor with the School of Electrical and Electronic Engineering. His research interests include multidisciplinary physics and electronic engineering applied to complex systems, networks, game theory, energy policy, stochastics, and biophotonics.